\documentclass[aps, prl,10pt,twocolumn,superscriptaddress,nofootinbib]{revtex4-1}
\usepackage[dvipsnames]{xcolor}
\usepackage{mathrsfs, amssymb, amsmath}  
\usepackage{epsfig, cancel}
\usepackage{latexsym}
\usepackage{caption}
\usepackage{natbib, comment}
\usepackage{booktabs}
\usepackage{url, natbib}
\usepackage{dcolumn}
\usepackage{multirow}
\usepackage{color}
\usepackage{cancel}
\usepackage{soul}
\usepackage[normalem]{ulem}
\usepackage{amsfonts,amssymb,amsmath, txfonts}
\usepackage{graphicx,epsfig}
\usepackage{psfrag}
\usepackage{hyperref}
\hypersetup{colorlinks=true}
\usepackage{mathtools}
\usepackage{enumitem}
\usepackage{float}
\restylefloat{table}
\usepackage[dvipsnames]{xcolor}
\usepackage{xcolor}
\hypersetup{ linktoc=all,
    colorlinks, linkcolor={blue},
    citecolor={darkred}, urlcolor={darkgreen}
}

\usepackage{mathrsfs, amssymb, amsmath, mathtools} 
\usepackage{multirow}
\usepackage{amsfonts,amssymb,amsmath, txfonts}
\usepackage{graphicx,epsfig}
\usepackage{hyperref}
\usepackage{xcolor}
\usepackage{subcaption}
\usepackage{booktabs}
\graphicspath{{Images/}}
\newlength{\DefaultTabColSep}
\definecolor{rosy}{RGB}{230,235,252}
\definecolor{myframetitle}{RGB}{90,89,170}
\definecolor{myblocktitle}{RGB}{140,185,249}
\definecolor{mytitle}{RGB}{10,80,26}

\definecolor{darkgreen}{RGB}{27,130,45}
\definecolor{darkblue}{rgb}{0,0,0.3}
\definecolor{darkred}{rgb}{0.7,0,0}

\definecolor{light gray}{RGB}{220,220,220}
\definecolor{dark purple}{RGB}{108,0,217}
\definecolor{pink}{RGB}{190,20,100}
\definecolor{orang}{RGB}{193,63,0}
\definecolor{green}{RGB}{11,98,17}
\definecolor{darkpink}{RGB}{153,0,76}
\definecolor{bluegreen}{RGB}{0,102,102}
\definecolor{greenlagan}{RGB}{0,102,0}
\definecolor{redgreen}{RGB}{102,102,0}
\definecolor{Redgreen}{RGB}{153,76,0}
\definecolor{vividviolet}{rgb}{0.62, 0.0, 1.0}
\definecolor{amaranth}{rgb}{0.9, 0.17, 0.31}
\definecolor{palatinateblue}{rgb}{0.15, 0.23, 0.89}
\definecolor{brightpink}{rgb}{1.0, 0.0, 0.5}
\definecolor{cornflowerblue}{rgb}{0.39, 0.58, 0.93}
\definecolor{deepcarminepink}{rgb}{0.94, 0.19, 0.22}
\definecolor{radicalred}{rgb}{1.0, 0.21, 0.37}
\def\H0{{\text{H}\hspace*{-2.05mm}\text{H} 0\hspace*{-1.35mm}0\ }}

\def\be{\begin{equation}}
\def\ee{\end{equation}}
\def\beq{\begin{equation}}
\def\eeq{\end{equation}}
\def\bea{\begin{eqnarray}}
\def\eea{\end{eqnarray}}

\begin{document}

\title{A Forecast for Testing the Decreasing \texorpdfstring{$H_0$}{H0} Trend in Time-Delay Cosmography}

\author{Shane Brady}\email{shane.brady@research.atu.ie}
\affiliation{Atlantic Technological University, Ash Lane, Sligo F91 YW50, Ireland}
\author{Eoin \'O Colg\'ain}\email{eoin.ocolgain@atu.ie}
\affiliation{Atlantic Technological University, Ash Lane, Sligo F91 YW50, Ireland}
\author{M. M. Sheikh-Jabbari} \email{jabbari@theory.ipm.ac.ir}
\affiliation{School of Physics, Institute for Research in Fundamental Sciences (IPM), P.O.Box 19395-5531, Tehran, Iran}

\begin{abstract}
In 2019 the H0LiCOW/TDCOSMO collaboration reported a decreasing Hubble constant $H_0$ trend with lens redshift in the flat $\Lambda$CDM model at low statistical significance ($< 2 \sigma$). The signal is expected if physics is missing from the $\Lambda$CDM model in the late universe. In the wake of recent TDCOSMO 2026 results, we revisit the feature. Although the statistical significance of the trend has increased in a like-for-like comparison with earlier studies, propagating mass-sheet transformation (MST) uncertainties reduces the significance to $\sim 1 \sigma$. As a result, Akaike Information Criterion (AIC) currently favours a constant $H_0$ over a decreasing $H_0$ in the ratio 2.4:1. Nevertheless, we highlight good agreement on a negative slope with independent observables and forecast that approximately 40 and 90 lenses will be required to substantiate the trend at  $\sim 2 \sigma$ and $\sim 3 \sigma$ significance, respectively, if the feature is physical and current uncertainties are representative. Even now, it is evident that the TDCOSMO 2026 $H_0$ constraint $H_0 \approx 71$ km/s/Mpc reflects an average over the lens redshift distribution and is naturally anchored at intermediate lens redshifts.   

\end{abstract}

\maketitle

\section{Introduction}
In response to cosmological tensions \cite{CosmoVerseNetwork:2025alb}, one can in principle prove that the flat $\Lambda$CDM model has broken down and also identify redshift ranges with missing physics \cite{Akarsu:2024qiq}. The central idea is that time-dependent model fitting parameters confronted to timeseries data is a hallmark of a model that fails to make unique predictions, i. e. a broken model. A simple analogy is a linear model confronted piecewise to quadratic data leading to evolution in the line intercept and slope. Applying the idea to cosmology, where redshift is a proxy for time, one studies the redshift dependence of $\Lambda$CDM parameters \cite{Krishnan:2020vaf, Krishnan:2022fzz} (see \cite{Vagnozzi:2023nrq, Akarsu:2024qiq} for reviews). Since observational systematics can mimic parameter evolution in a single observable, to overcome systematics one must demonstrate qualitatively similar evolution in independent physical observables. 

The inspiration for this line of research goes back to $H_0$ tension \cite{Planck:2018vyg, Riess:2021jrx} and time-delay cosmography \cite{H0LiCOW:2019pvv, Millon:2019slk}. In 2019, H0LiCOW/TDCOSMO collaborations reported a decreasing $H_0$ trend with lens redshift with statistical significance $\sim 1.9 \sigma$ \cite{H0LiCOW:2019pvv} and $\sim 1.7 \sigma$ \cite{Millon:2019slk} assuming a fixed mass-sheet transformation (MST) parameter $\lambda_{\mathrm{int}}= 1$. The trend is not due to an obvious systematic \cite{Millon:2019slk} and there are further reasons to believe it is physical.

First, independent groups have reported qualitatively similar $H_0$ evolution with a negative slope in independent observables \cite{Krishnan:2020obg, Dainotti:2021pqg, Dainotti:2022bzg, Jia:2022ycc, Dainotti:2023yrk, Malekjani:2023ple, Jia:2024wix, Xu:2024xgi, Lopez-Hernandez:2024osv, Tang:2024nhd, Kalita:2025jqz, Yang:2026vti, Dainotti:2026fsm}. Secondly, the Dark Energy Spectroscopic Instrument (DESI) currently reports a dynamical dark energy (DE) signal \cite{DESI:2024mwx, DESI:2024hhd, DESI:2024jxi, DESI:2025zgx}, which can be mapped back into a redshift-dependent matter density parameter $\Omega_m$ in the $\Lambda$CDM model \cite{Colgain:2024mtg}. Conceptually, there cannot be a deviation from $\Lambda$CDM behaviour in the late universe unless tomographically $\Omega_m$ shows evolution across redshift bins. Currently, it is unclear if the $\Omega_m$ trend is monotonic or not, but the increasing $\Omega_m$ trend noted in 2022 \cite{Colgain:2022nlb, Colgain:2022rxy} is expected if $H_0$ decreases, since the parameters are negatively correlated; see \cite{Colgain:2024ksa} for explicit confirmation. What makes time-delay cosmography interesting is that it constrains $H_0$ better than $\Omega_m$, so even if a negative correlation between $H_0$ and $\Omega_m$ explains the trend, its influence in time-delay cosmography is weaker. 

In this letter we revisit earlier H0LiCOW/TDCOSMO observations \cite{H0LiCOW:2019pvv, Millon:2019slk} with the most recent TDCOSMO 2026 eleven-lens sample \cite{TDCOSMO:2026dxn}, an extension of an earlier eight-lens sample \cite{TDCOSMO:2025dmr}. Our objective is to understand whether the decreasing $H_0$ trend persists under similar assumptions to earlier studies \cite{H0LiCOW:2019pvv, Millon:2019slk} and determine the extent to which the statistical significance of the signal is washed out once uncertainties in the MST parameter are propagated. We also document the constant versus decreasing $H_0$ preference in the TDCOSMO 2026 sample and forecast the number of lenses required to arrive at $\sim 2 \sigma$ and $\sim 3 \sigma$ observations if the trend proves to be physical. See \cite{Schiavone:2022wvq, Montani:2023xpd, Montani:2023ywn, Schiavone:2024heb, Montani:2024ntj, Dai:2026pvx} for musings on the physical interpretation of a redshift-dependent $\Lambda$CDM parameter $H_0$.

\section{Background}

In time-delay cosmography, the excess time delay for a lensed image relative to an unlensed image is \cite{Refsdal:1964blz, Shapiro:1964uw} 
\begin{equation}
t (\theta, \beta) = \frac{D_{\Delta t}}{c} \left[ \frac{(\theta-\beta)^2}{2}- \psi(\theta) \right],
\end{equation}
where $\theta$ is the image position, $\beta$ is the source position, $\psi(\theta)$ is the deflection potential and the time-delay distance is
\begin{equation}
    D_{\Delta t} := (1+z_{\mathrm{d}}) \frac{D_{\mathrm{d}} D_{\mathrm{s}}}{D_{\mathrm{ds}}}, 
\end{equation}
where $z_\mathrm{d}$ is the lens redshift and $D_{\mathrm{d}}$, $D_{\mathrm{s}}$ and $D_{\mathrm{ds}}$ denote angular diameter distances from the observer to the lens, the observer to the source, and the lens to the source, respectively. Since each angular distance is inversely proportional to $H_0$, one has $D_{\Delta t} \propto H_0^{-1}$.

\begin{table*}[t]
\centering
\renewcommand{\arraystretch}{1.3}
\begin{tabular}{|c|c|c|c|c|c|c|}
\hline
Lens & $z_\mathrm{d}$  & $H_0 (\lambda_{\mathrm{int}}=1)$ (km/s/Mpc) & $\lambda_{\mathrm{int}}$ & $H_0 \equiv H_0 (\lambda_{\mathrm{int}}=1) \, \lambda_{\mathrm{int}}$ (km/s/Mpc) & Ref. \\
\hline\hline
WGD 2038–4008  & $0.228$ & $65^{+23}_{-14}$ & $1$ & $65^{+24}_{-15}$ & \cite{TDCOSMO:2024rwr} \\
RX J1131--1231 (H0LiCOW) & $0.295$ & $78.2 \pm 3.4$ & $1$ & $78.2 \pm 7.8$ & \cite{H0LiCOW:2019pvv} \\
RX J1131--1231 (TDCOSMO) & $0.295$ & - & Free & $77.1^{+7.3}_{-7.1}$ & \cite{TDCOSMO:2023hni} \\
PG 1115+080 & $0.310$ & $81.1^{+8.0}_{-7.1}$ & $1$ & $81.1^{+10.8}_{-10.2}$ & \cite{H0LiCOW:2019pvv}\\
SDSS J1433+6007 & $0.407$ & - & $1.12^{+0.05}_{-0.06}$& $73.2^{+4.8}_{-4.7}$ & \cite{TDCOSMO:2026qso} \\
HE 0435--1223 & $0.455$ & $71.7^{+4.8}_{-4.5}$ & $1$ & $71.7^{+8.0}_{-7.9}$ & \cite{H0LiCOW:2019pvv} \\
J1537--3010 & $0.590$ & - & $0.89^{+0.09}_{-0.07}$ & $75.5^{+9.3}_{-5.8}$ & \cite{Galan:2026mon} \\
DES J0408--5354 & $0.597$ & $74.2^{+2.7}_{-3.0}$& $1$ & $74.2^{+7.2}_{-7.3}$ & \cite{DES:2019fny} \\
B1608+656 & $0.630$ & $71.0^{+2.9}_{-3.3}$ & $1$ & $71.0^{+7.0}_{-7.2}$ & \cite{H0LiCOW:2019pvv} \\
WFI 2033--4723 & $0.657$ & $71.6^{+3.8}_{-4.9}$ & $1$ & $71.6^{+7.5}_{-8.1}$ & \cite{H0LiCOW:2019pvv} \\
HE 1104--1805 & $0.729$ & $64.2^{+5.8}_{-5.0}$ & $1$ & $64.2^{+8.2}_{-7.6}$ & \cite{TDCOSMO:2025ruf} \\
%RX J1131--1231 (H0LiCOW) & $0.295$ & $78.2 \pm 3.4$ & 1 & \\
SDSS J1206+4332 & $0.745$ & $68.9^{+5.4}_{-5.1}$ & $1$ & $68.9^{+8.2}_{-8.0}$ & \cite{H0LiCOW:2019pvv}\\
\hline
\end{tabular}
\caption{Lens $H_0$ determinations ordered by increasing lens redshift $z_\mathrm{d}$.}
\label{tab:h0_constraints}
\end{table*}

Nevertheless, there is a complication. The mass-sheet degeneracy (MSD) \cite{Falco:1985, Schneider:2013sxa} represents the main systematic uncertainty that changes the time-delay distance by a multiplicative factor $D_{\Delta t} \rightarrow \lambda^{-1} D_{\Delta t}$, where $\lambda$ is the MST parameter. Note, this implies $H_0 \rightarrow \lambda H_0$, so $H_0$ is defined up to a factor of the MST parameter. The MST parameter is a product 
\begin{equation}
    \label{eq:lambda}
    \lambda := \lambda_{\mathrm{int}} (1-\kappa_{\textrm{ext}}), 
\end{equation}
where $\lambda_{\mathrm{int}}$ encodes a transformation of the primary deflector’s mass profile, and $\kappa_{\mathrm{ext}}$ denotes the convergence due to matter along the line of sight. TDCOSMO analysis places $\lambda_{\mathrm{int}}$ at a $\sim 10 \%$ uncertainty \cite{Galan:2026mon, TDCOSMO:2026qso, TDCOSMO:2026dxn}, whereas $\kappa_{\mathrm{ext}}$ represents a smaller $\lesssim 3 \%$ uncertainty \cite{Johnson:2026mgk} once $\lambda_{\mathrm{int}}$ is fitted. Since  $\kappa_{\rm{ext}}$ is a subleading uncertainty relative to $\lambda_{\mathrm{int}}$, we do not consider it in our analysis. As noted in \cite{TDCOSMO:2024rwr}, $\Omega_m$ is very weakly constrained by time-delay cosmography applied to a single lens \cite{H0LiCOW:2016tzl, H0LiCOW:2019pvv}, so whether one fixes it to a canonical value $\Omega_m \sim 0.3$, or marginalises over it, the $H_0$ determination from a single lens does not vary much.

The results in Table \ref{tab:h0_constraints} come from the literature. In nine lenses, the MST parameter has been fixed to $\lambda_{\mathrm{int}} = 1$. Indeed,   the result in \cite{TDCOSMO:2025ruf} is quoted as $H_0 = 64.2^{+5.8}_{-5.0} \,  \lambda_{\mathrm{int}}$ km/s/Mpc, which makes it explicit that $H_0$ is only defined up to the multiplicative factor $\lambda_{\mathrm{int}}$. In contrast, the results for RX J1131--1231, SDSS J1433+6007 and J1537--3010 incorporte a fitted $\lambda_{\mathrm{int}}$ even if the $\lambda_{\mathrm{int}}$ constraint is not explicitly quoted. We can compare the RX J1131--1231 result with the fitted $\lambda_{\mathrm{int}}$, $H_0 = 77.1^{+7.3}_{-7.1}$ km/s/Mpc \cite{TDCOSMO:2023hni} to the earlier finding $H_0= 78.2 \pm 3.4$ km/s/Mpc \cite{H0LiCOW:2019pvv} with $\lambda_{\mathrm{int}} = 1$. Evidently, while the shift in the central value need not be great, the $H_0$ errors can double when one relaxes $\lambda_{\mathrm{int}} = 1$. 

Here, we bypass the absence of explicit lens-by-lens $\lambda_{\mathrm{int}}$ results in the literature by propagating an error for $\lambda_{\mathrm{int}}$ to $\lambda_{\mathrm{int}}=1$ lenses. The value we use is the intrinsic scatter across the lenses in TDCOSMO 2026 sample, $\sigma( \lambda_{\mathrm{int}}) = 0.093^{+0.038}_{-0.026}$ \cite{TDCOSMO:2026dxn}. Concretely, we assume $\lambda_{\mathrm{int}}$ is normally distributed about $\lambda_{\mathrm{int}} = 1$, $\lambda_{\mathrm{int}} \sim \mathcal{N} (1, \sigma_{\lambda}^2)$ with $\sigma_{\lambda} = 0.09$. We define $H_0 \equiv H_0 (\lambda_{\mathrm{int}}=1) \,  \lambda_{\mathrm{int}}$, so that standard Gaussian error propogation dictates 
\[
\sigma_{H_0} = \sqrt{\sigma^2_{H_0(\lambda_{\mathrm{int}}=1)} + H^2_0(\lambda_{\mathrm{int}}=1) \, \sigma^2_{\lambda}}\,.
\] 
Applying this to antisymmetric $H_0 (\lambda_{\mathrm{int}=1})$ errors for the $\lambda_{\mathrm{int}} = 1$ lenses in Table \ref{tab:h0_constraints} we arrive at the errors in the fifth column. It is also interesting to apply this to the historical RX J1131-1231 result, where $H_0 (\lambda_{\mathrm{int}}=1)= 78.2 \pm 3.4$ km/s/Mpc gets converted to $H_0 = 78.2 \pm 7.8$ km/s/Mpc, which is a more conservative error than the value from the later paper with fitted $\lambda_{\mathrm{int}}$. Evidently, our MST error propagation need not be underestimating errors. 

We remark that fitting $\lambda_{\mathrm{int}}$ for each lens also shifts the central value. Given the scatter in $\lambda_{\mathrm{int}}$, 10\% shifts are plausible, but as we see in the case of RX J1131-1282, relaxing fixed $\lambda_{\mathrm{int}}=1$ may simply inflate errors without leading to a pronounced shift in the central value. {Nevertheless, as is clear from Table \ref{tab:h0_constraints}, $\lambda_{\mathrm{int}}$ is displaced by $\sim 10\%$ for SDSS J1433+6007 and J1537--3010, which naively shifts $H_0 \approx 73 \mathrm{ km/s/Mpc } \rightarrow \, \approx 65$ km/s/Mpc and $H_0 \approx 75 \mathrm{ km/s/Mpc } \rightarrow \, \approx 84$ km/s/Mpc  respectively, once one imposes $\lambda_{\textrm{int}}=1$.\footnote{It is worth noting that larger values of $H_0 \gtrsim 73$ km/s/Mpc are disfavoured by $\sim 2 \sigma$ by the age of globular clusters and astrophysical objects \cite{Krishnan:2021dyb, Vagnozzi:2021tjv}.} In the appendix of \cite{TDCOSMO:2026qso}, this shift is studied in greater detail, where it is noted that $\kappa_{\mathrm{ext}}$ compensates $\lambda_{\mathrm{int}}$ to some extent so that the shift is lessened. Since $\lambda$ in  equation (\ref{eq:lambda}) is a product of $\lambda_{\mathrm{int}}$ and $\kappa_{\mathrm{ext}}$, this behaviour is expected.}
While a full lens-by-lens analysis would be desirable, given results in the literature, we do not expect the qualitative conclusions of the present study to depend sensitively on the precise treatment of individual lenses.

\section{Decreasing $H_0$}
As is clear from the Hubble constant values in Table \ref{tab:h0_constraints}, there is a decreasing trend of $H_0$ with lens redshift $z_{\mathrm{d}}$. Previously, both H0LiCOW and TDCOSMO collaborations commented on this trend in a six-lens \cite{H0LiCOW:2019pvv} and seven-lens system \cite{Millon:2019slk}, reporting a statistical significance of $\sim 1.9 \sigma$ and $\sim 1.7 \sigma$, respectively. This motivates the first exercise, where we fit a line model 
\begin{equation}
\label{eq:line}
    H_0(z) = m \, z_\mathrm{d} + c, 
\end{equation}
to the nine $H_0 ({\lambda_{\mathrm{int}}=1})$ values in Table \ref{tab:h0_constraints}. This constitutes a like-for-like comparison with fixed $\lambda_{\mathrm{int}}=1$ to earlier analyses \cite{H0LiCOW:2019pvv, Millon:2019slk}. We make use of Markov chain Monte Carlo (MCMC) through \textit{emcee} \cite{Foreman-Mackey:2012any} and incorporate antisymmetric errors by identifying if the line is above or below the central values and selecting the appropriate error to define the log-likelihood for each $(m,c)$-pair in the MCMC chain. The resulting slope for the nine lenses subject to the $\lambda_{\mathrm{int}} = 1$ assumption is 
\begin{equation}
    \label{eq:slope1}
    m = -20.6^{+9.4}_{-9.5}\,, 
\end{equation}
which corresponds to a $\sim 2.2 \sigma$ effect. The slope shows remarkable agreement with the descending $H_0$ trend in independent binned data comprising megamasers, cosmic chronometers, SNe Ia and BAO, $m = -21.7 \pm 9.4$  \cite{Krishnan:2020obg}, which corresponds to a naive $\sim 2.3 \sigma$ statistical significance. However, using simulations following H0LiCOW analysis \cite{H0LiCOW:2019pvv}, the statistical significance was independently estimated at $\sim 2.1 \sigma$ \cite{Krishnan:2020obg}. We attribute the small difference to the use of \texttt{scipy.optimize.curve\_fit} in place of MCMC and inflate the error accordingly to $m = -21.7 \pm 10.3$ to bring the statistical significances into line at $\sim 2.1 \sigma$. We note that  as  demonstrated in \cite{Akarsu:2024hsu}, simulations and MCMC for line fits exhibit excellent agreement,  negating the need to run simulations here.

Neglecting the MST uncertainty, we have two largely independent and Gaussian constraints on the same quantity, namely the slope of $H_0(z)$. Under these assumptions, one can symmetrise the errors and combine them using inverse-variance weighting  
\begin{equation}
\bar{m} = \frac{\sum_i m_i/\sigma_i^2}{\sum_i 1/\sigma_i^2}, \quad \bar{\sigma} = \left(\sum_i \frac{1}{\sigma^2_i}\right)^{-\frac{1}{2}}\,. 
\end{equation}
We find $\bar{m} = -21.2 \pm 6.9$, which is just over $3 \sigma$ removed from the horizontal. The combination illustrates how two largely independent $\sim 2 \sigma$ measurements of the same quantity can jointly approach the $\sim 3 \sigma$ threshold.

Nevertheless, this is not the whole story, we have omitted the MST uncertainty that inflates the error. Repeating the analysis with the 11 lenses with the propagated $\lambda_{\mathrm{int}}$ errors from the fifth column in Table \ref{tab:h0_constraints}, we find 
\begin{equation}
    \label{eq:slope2}
    m = - 16 \pm 15, 
\end{equation}
which corresponds to a statistical significance of just over $1 \sigma$. Combining this with $m = -21.7 \pm 10.3$, we get $\bar{m} = -19.9 \pm 8.5$, a statistical significance of $ \sim 2.3 \sigma$. 

\begin{figure}[t]
\centering
\includegraphics[scale=0.4]{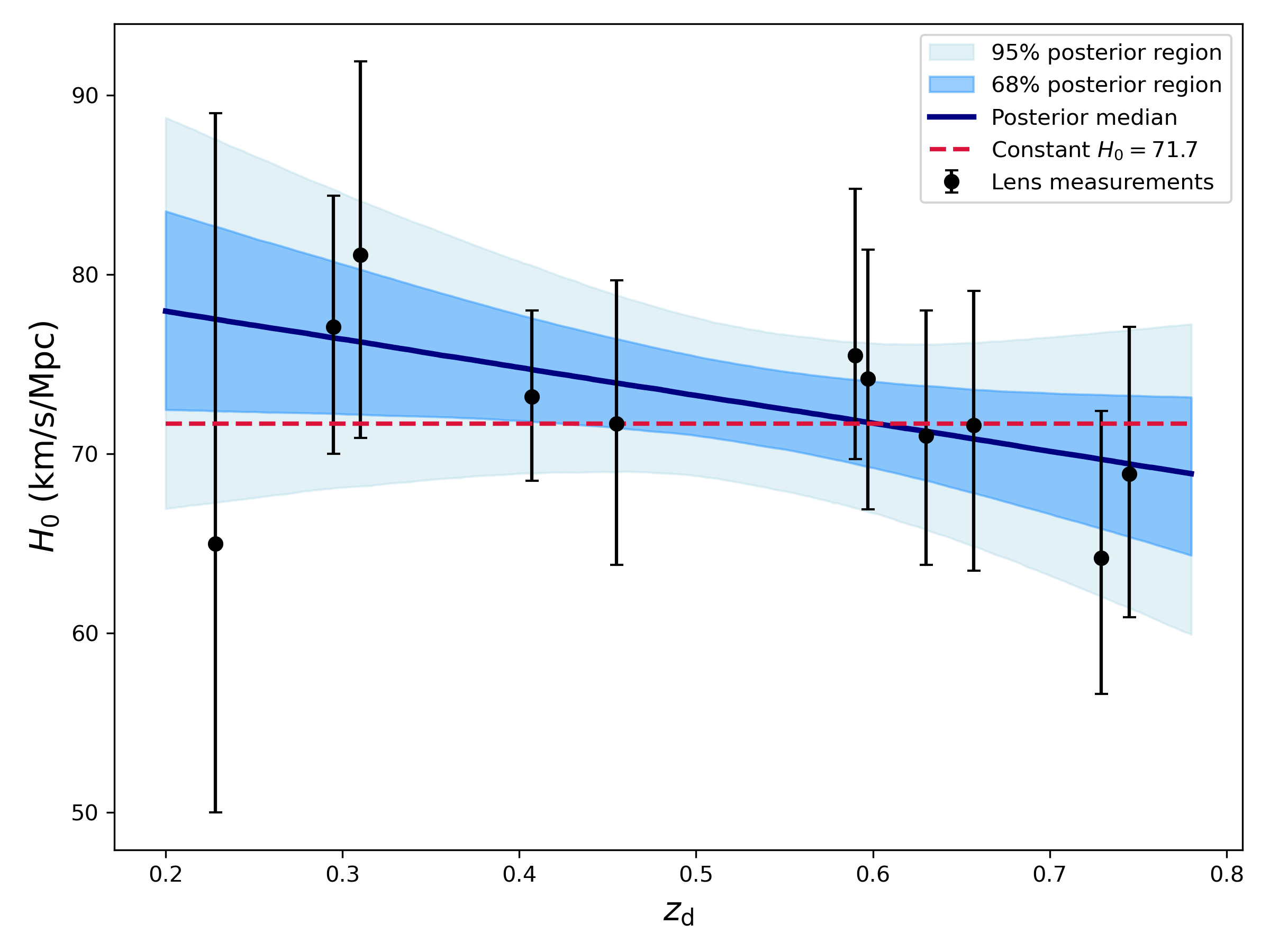}
\caption{MST uncertainty inflated $H_0$ determinations alongside best fit decreasing $H_0(z) = m \, z_{\mathrm{d}} + c$ with credible intervals. For comparison, we include best fit constant $H_0$.}
\label{fig:H0}
\end{figure}

In Fig. \ref{fig:H0} we plot the preferred line with credible intervals alongside $H_0$ constraints with propagated MST uncertainties. There are a number of interesting features worth commenting upon. First, despite the $\gtrsim 10 \%$ errors, there is a noticeable decreasing trend. Secondly, given that the SH0ES \cite{Riess:2021jrx} and H0DN collaborations \cite{H0DN:2025lyy} determine $H_0$ in the redshift range $z \lesssim 0.15$ to be $H_0 \sim 73$ km/s/Mpc with percent level errors, the most local lens would be in mild tension if not for the $\approx 35 \%$ error bar. As a result, the central value of this data point not only deviates from local $H_0$ expectations, but also from the conjectured decreasing $H_0$ trend. Thirdly, the decreasing trend is driven by two lenses at $z_{\mathrm{d}} \approx 0.3$ and the two at $z_{\mathrm{d}} \approx 0.75$ spanning three independent and blinded studies with different assumptions \cite{H0LiCOW:2019pvv, TDCOSMO:2023hni, TDCOSMO:2025ruf}. Thus, one cannot easily argue that the signal is driven by a single outlying lens. Finally, if the decreasing $H_0$ trend is physical, it should be clear that constant $H_0$ in the eleven-lens sample measures an average $H_0$ over the lens redshift distribution rather than a local $H_0$. Consequently, the inferred average $H_0 \approx 71.7$ km/s/Mpc is naturally anchored at intermediate redshifts, where the sample is concentrated. The concern here is that the $H_0$ value may depend on the redshift distribution of the sample itself. 

At this juncture, we should confirm that constant $H_0$ is preferred once MST uncertainties are propagated. To do so, we employ the Akaike Information Criterion (AIC) corrected for small samples: 
\begin{equation}
    \mathrm{AIC}_c = \frac{2 k N}{N-k-1} - 2 \ln \mathcal{L}_{\mathrm{max}}, 
\end{equation}
where $k \in \{1, 2\}$ is the number of free parameters, $N=11$ is the size of the sample and $\mathcal{L}_{\mathrm{max}}$ is the likelihood maximum. We find $\mathrm{AIC}^{\mathrm{const}}_c = 71.7$ for constant $H_0$ and $\mathrm{AIC}^{\mathrm{line}}_c = 73.5$ for the line, which translates into a preference for constant $H_0$. Further defining $\Delta_i = \mathrm{AIC}_{c, \, i}- \mathrm{AIC}_{c \, \mathrm{min}}$, we can calculate Akaike weights: 
\begin{equation}
    w_i = \frac{\exp (-\Delta_i/2)}{\sum_j \exp (-\Delta_j/2)},  
\end{equation}
which leads to $w_{\mathrm{const}} = 0.71$ and $w_{\mathrm{line}} = 0.29$, corresponding to the $\mathrm{AIC}_c$ favouring constant $H_0$ over decreasing $H_0$ in the ratio 2.4:1. 

\section{Forecast}
In this section, we assume that the decreasing $H_0$ trend with slope $m = -16 \pm 15$ is physical, namely due to missing physics, and forecast the number of lenses $N$ required to push the current $\sim 1 \sigma$ finding to a $ \sim 2 \sigma$ and $\sim 3 \sigma$ result. Before running a Monte Carlo, we comment on the theoretical expectation. Assuming a Gaussian error on the slope $\sigma_{m}$, the statistical significance of a non-zero slope is 
\begin{equation}
    Z \equiv |m|/\sigma_m.
\end{equation} 
Here we inject the slope, so the statistical significance of $Z$ is dictated by $\sigma_{m}$, which is expected to scale with the sample size, $\sigma_{m} \propto N^{-\frac{1}{2}}$, thus implying $Z(N) \propto N^{\frac{1}{2}}$. Given that $Z (N \approx 10) \sim 1 \sigma$, one anticipates that $Z(N \approx 40) \sim 2 \sigma$ and $Z(N \approx 90) \sim 3 \sigma$. We now run a Monte Carlo simulation to confirm this expectation. 

To do so, we extract the median slope $m = -15.9$ and intercept $c = 81.2$ from the MCMC chain. We assume $N$ lenses in the range $11 \leq N \leq 150$. For each $N$ we generate $M = 10^4$ mock realisations of the data by selecting $N$ lens redshifts $z_{\mathrm{d}}$ randomly and uniformly in the range $z_{\mathrm{d}} \in [0.228, 0.745 ]$. We symmetrise the $H_0$ errors for the observed data in Table \ref{tab:h0_constraints} and draw $N$ $H_0$ errors randomly from the set of observed errors. For each mock we define $H_0$ central values on the line (\ref{eq:line}) before using the error to define a displaced $H_0$ value $H_0 \rightarrow H_0 + \mathcal{N}(0, \sigma^2_{H_0})$. For each of the $M$ mocks, we refit a line using generalised least squares and collect $M$ slopes $m$, before extracting a median negative slope $m_{50}$, an 84$^{\mathrm{th}}$ percentile slope $m_{84}$ and calculating the statistical significance of the negative slope $Z(N) \equiv |m_{50}|/(m_{84}-m_{50})$.

\begin{figure}[t]
\centering
\includegraphics[scale=0.4]{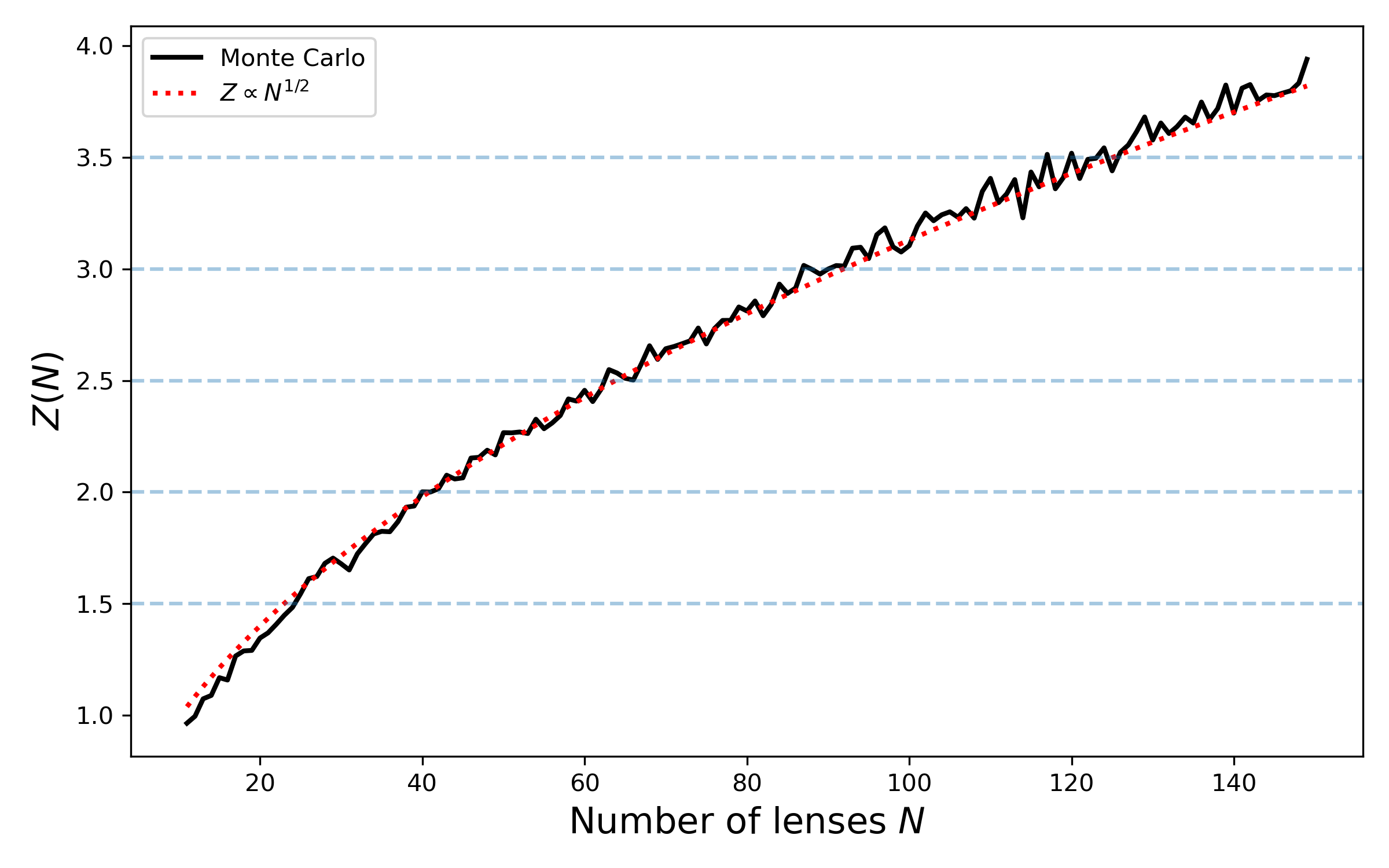}
\caption{The statistical significance of the decreasing slope $Z(N)$ with increasing $N$.}
\label{fig:Z}
\end{figure}

We emphasise that we have fixed the slope. One could generalise our mocks by picking $m$ from the MCMC chain or alternatively in a normal distribution $\mathcal{N}(-16, 15^2)$, but this is guaranteed to recover $m = 0$ at $\sim 1 \sigma$ irrespective of the number of lenses $N$ in the mock. One learns little from such an exercise as the outcome is fixed by the slope and error in  (\ref{eq:slope2}).

In Fig. \ref{fig:Z} we show the statistical significance of the decreasing $H_0$ trend for increasing number of lenses $N$ alongside the theortical expectation $Z \propto N^{\frac{1}{2}}$. By plotting $(m_{84}-m_{50}) \sqrt{N}$ we recover the expected constant behaviour at $N \gtrsim 30$, but there is a small sample correction so that $Z \propto N^{0.53}$ over the range $11 \leq N \leq 150$. From Fig. \ref{fig:Z} we confirm that our earlier back of the envelope calculation was correct and approximately 40 and 90 lenses are required for a $\sim 2 \sigma$ and $\sim 3 \sigma$ result. 

It is worth stressing here that this forecast is conservative, since it assumes that the $H_0$ errors in the fifth column of Table \ref{tab:h0_constraints} will be representative of future lenses. The reality here is that JWST, ELT, Euclid and Romans Space Telescope will lead to better stellar kinematics and higher quality lens imaging, while more uniform analysis pipelines are also expected to reduce scatter.

\section{Conclusions}

The main takeaway from this letter is that a decreasing $H_0$ trend in time-delay cosmography reported originally in \cite{H0LiCOW:2019pvv, Millon:2019slk} persists at low statistical significance in a larger lens sample. More precisely, we found that the statistical significance of the negative slope increased to $\sim 2.2 \sigma$ in a like-for-like comparison with fixed $\lambda_{\mathrm{int}}$, but is diluted to $\sim 1 \sigma$ once MST uncertainties are propagated. The preferred slope nevertheless remains negative and is in good agreement with independent determinations based on different observables \cite{Krishnan:2020obg,Dainotti:2021pqg,Dainotti:2022bzg,Jia:2022ycc,Dainotti:2023yrk,Malekjani:2023ple,Jia:2024wix,Xu:2024xgi,Lopez-Hernandez:2024osv,Tang:2024nhd,Kalita:2025jqz,Yang:2026vti,Dainotti:2026fsm}. In particular, combining the MST-inflated lensing result with the $\sim2.1\sigma$ negative slope reported in \cite{Krishnan:2020obg} yields $\bar m=-19.9\pm8.5$, corresponding to a nominal significance of $\sim 2.3\sigma$.

Given the low significance of the trend in lensing data alone, we confirmed that the TDCOSMO assumption of a constant $H_0$ is currently preferred by AIC. Nevertheless, if the decreasing trend is physical, a constant $H_0$ analysis effectively measures a weighted average over the lens redshift distribution. Consequently, the TDCOSMO 2026 determination, $H_0 \approx 71$ km/s/Mpc \cite{TDCOSMO:2026dxn}, is naturally anchored to lenses at intermediate redshifts rather than the most local systems. This can be contrasted with SH0ES/H0DN determinations \cite{Riess:2021jrx, H0DN:2025lyy} at more local redshifts. 

Finally, we provided a forecast confirming analytic expectations that approximately 40 and 90 lenses will be required to confirm the decreasing $H_0$ trend at $\sim 2 \sigma$ and $\sim 3 \sigma$ significance, respectively, assuming current errors. As emphasised by the TDCOSMO collaboration, doubly lensed quasars are far more abundant than quadruply lensed quasers, and they promise to increase time-delay cosmography samples by a factor of five \cite{TDCOSMO:2025ruf}. In addition, Euclid is expected to discover in excess of $10^5$ strong lenses during its six-year mission \cite{Lines:2025}. This suggests that the principal limitation will not be the lenses, but the acquisition of time-delays, kinematic and environmental information. In terms of a timeline, TDCOSMO added 3 lenses between 2025 \cite{TDCOSMO:2025dmr} and 2026 \cite{TDCOSMO:2026dxn}, suggesting that we may arrive at a 40 lens sample in the next decade, and potentially a $\sim 2 \sigma$ result, if the current pace of sample growth continues.

\section*{Acknowledgements} 
We thank Tao Yang for comments on the manuscript. The work of MMShJ is in part supported by the INSF research chair grant No.4045163. 

\bibliography{refs}

\end{document}